# Magnetic phase diagrams in the *H-T* plane of the magnetically strongest sigma-phase Fe-V compounds

M. Sikora[1] and S. M. Dubiel[2*]

[1]AGH University of Science and Technology, Academic Centre for Materials and Nanotechnology, al. A. Mickiewicza 30, 30-059 Kraków, Poland, [2]AGH University of Science and Technology, Faculty of Physics and Applied Computer Science, al. A. Mickiewicza 30, 30-059 Kraków, Poland

Magnetization measurements were performed on two sigma-phase samples of $Fe_{100-x}V_x$ ($x$=35.5, 34.1) vs. temperature, T, and in DC magnetic field, of various amplitudes. Using three characteristic temperatures, magnetic phase diagrams in the H-T plane have been designed testifying to a re-entrant character of magnetism. The ground magnetic state, a spin glass (SG), was evidenced to be composed of two sub phases: one with a weak irreversibility and the other with a strong irreversibility. Two critical lines were reconstructed within the SG state. Both of them show a crossover from the Gabay-Toulouse behavior (low field) to a linear and/or quasi-Almeida-Touless behavior. A strong difference in the effect of the applied magnetic field on the SG phase in the two samples was revealed.

An interest in spin-glasses (SGs) had is apogee some 3-4 decades ago. Despite a huge number of papers have been published – for a review see [1-4] – a full understanding of the SGs and their behavior(s) has not been achieved yet. One of the main reasons for this situation follows, in our opinion, from the fact that phenomena originally observed in systems defined as SG (recently known rather as canonical SGs, C-SGs) were also revealed in systems that cannot be classified as the C-SGs because a concentration of magnetic atoms in these systems is significantly higher than the so-called percolation limit (the systems are termed as concentrated or cluster SGs). Concerning the latter, as the only criterion for the C-SGs, one usually considers a chemical content of the magnetic entities. Whereas, the magnetic interactions i.e. their strength and range seem to be even more important factors determining whether or not a system shows a spin-glass like behavior. The latter plus a thermal history dependent actual degree of alloys homogeneity make, on one hand, the percolation limit unpredictable, and, on the other hand, a diversity of systems exhibiting the *SG*-like features is huge and difficult to be described in a unified way. The importance of the range of magnetic interactions can be well illustrated with σ-phase Fe-X (X=Cr, Mo, Re, V) alloys the magnetism of which is highly itinerant [5,6], hence the range of the magnetic interactions is very long (in this aspect they fulfill the mean-field theory assumption concerning the SGs). These systems, despite a very high content of Fe (~50 at %), show features characteristic of the C-SGs [7-9]. In particular, a well-defined cusp in the ac-susceptibility hardly depends on the frequency.

One of open questions relevant to the SGs is a dependence of characteristic temperature(s), $T_{SG}$, on an external magnetic field, $H$. In the mean-field theory (MFT), the $T_{SG}$-$H$ relationship reads as follows:

$$\left(1 - \frac{T}{T_{SG}}\right) = aH^\varphi \qquad (1)$$

Three different predictions concerning the value of $\varphi$ and thus the $T_{SG}$-$H$ relationships are identified throughout the literature: (1) $\varphi$=2/3 known as the Almeida-Thouless (*AT*) line [10], (2) $\varphi$ =2 known as the Gabay-Toulouse (*GT*) line [11], and (3) $\varphi$ =1 [12]. Examples of the *AT*-line are numerous e. g. [13,14], those of the *GT*-line are more rare e. g. [15,16], and the line with $\varphi$ =1 was reported twice [12,17]. The MFT predicts two critical lines in Heisenberg SGs. Also two lines should exist in the so-called re-entrant SGs [11]. Usually a co-existence of the *AT* and *GT* lines was reported e. g. [18, 19]. For EuSrS, an SG insulator, a cross-over from the GT-like to the AT-like behavior was observed [20].

In this paper is reported a clear-cut evidence on the co-existence of two lines reveled with dc-magnetic measurements in two σ-FeV alloys. Interestingly, the character of the two lines changes with *H* from *AT*- and *GT*-like into $\varphi$ =1-like for higher *H* – values.

The magnetic measurements were carried out with a LakeShore type 7407 vibrating sample magnetometer. Small samples were cut out from larger ingots and installed in the magnetometer in such a way that the demagnetization effects were minimized. The demagnetization factors were estimated from magnetization, *M*, curves measured at low fields, *H*. The obtained values were checked by describing the samples as rectangular prisms and compared to the results provided by equations given in [21]. The estimated values for the demagnetization factors were $\eta$ = 0.125 and 0.15 for the $\sigma$-$Fe_{100-x}V_x$ samples of x=33.5 and x=34.1, respectively. Details on samples' preparation and characterization are given elsewhere [22,23]. The magnetization curves were recorded for both samples as functions of temperature upon zero field cooling (ZFC) and field cooling (FC) conditions, and in an external magnetic field, *H*, whose magnitudes were restricted to values below 500 Oe. Examples of the measured *M*-curves are presented in Fig. 1.

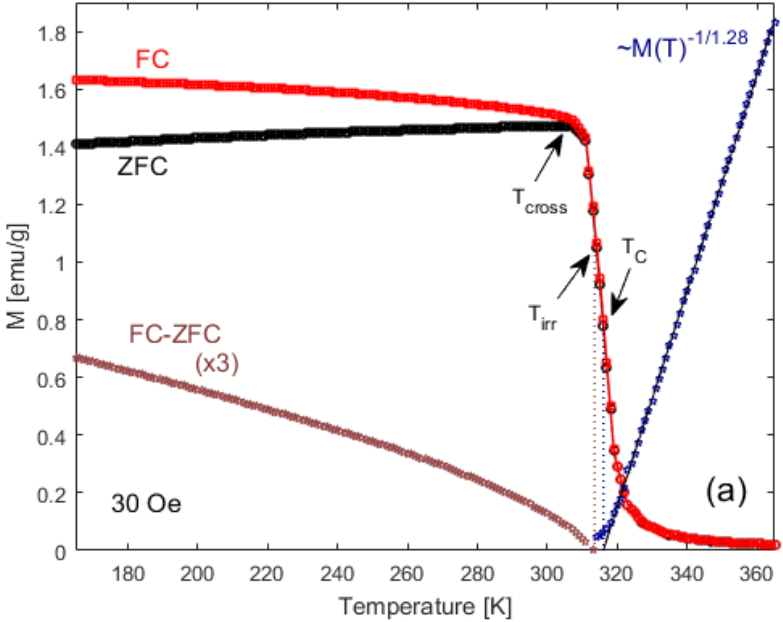

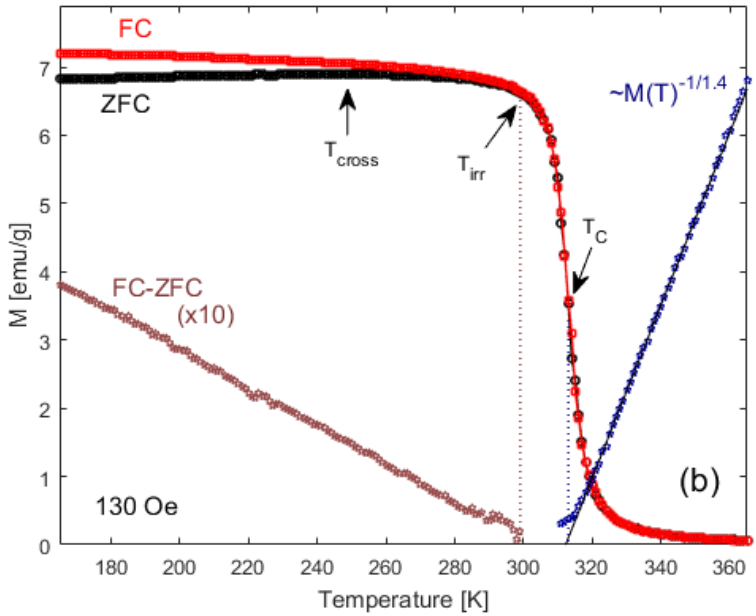

Fig. 1 Magnetization, M, curves measured vs. temperature in zero-field cooled (ZFC) and field cooled (FC) conditions on the σ-Fe$_{100-x}$V$_x$ samples: (a) for x=35.5, and (b) for x=34.1, in an external magnetic field shown. The susceptibility data measured in the paramagnetic range analyzed in terms of the Kouvel-Fisher procedure (solid line) are shown on the right-hand side of each panel.

Three characteristic temperatures derived from the curves were considered viz. (1) the magnetic ordering temperature, $T_C$, (2) the irreversibility temperature, $T_{irr}$, and (3) the crossover temperature, $T_{cross}$. The modified Kouvel-Fisher approach [24] applied to the susceptibility in the paramagnetic range yielded values of $T_C$ viz. 316 K for x=33.5, and 314 K for x=34.1. They compare well with those reported previously [20,21]. Furthermore, the critical exponent γ was determined from this procedure as 1.3 for x=33.5 and 1.4 for x=34.1. These values are slightly smaller than the value of 1.6 found previously for σ-FeV with x=48 and $T_C$= 42 [25]. The values of $T_{irr}$ were defined as temperature of bifurcation in the M-curves. $T_{irr}$ is regarded as a temperature at which a spin-freezing process starts and irreversibility phenomena occur at T < $T_{irr}$. The latter can be clearly seen by plotting a difference between the FC and ZFC M-curves. The corresponding data shown in Fig. 1 give a clear evidence that a degree of the irreversibility increases with a decrease of T, and that it is slightly weaker for x=33.5. Finally, $T_{cross}$-values were determined from the maximum of the ZFC-curves. This temperature is interpreted as the one marking a transition of system from a weak into a strong irreversibility of a spin-glass state. Using all values of these 3 characteristic temperatures obtained for different values of the magnetic field applied, a magnetic phase diagrams in the H-T plane have been constructed. They are displayed in Fig. 2.

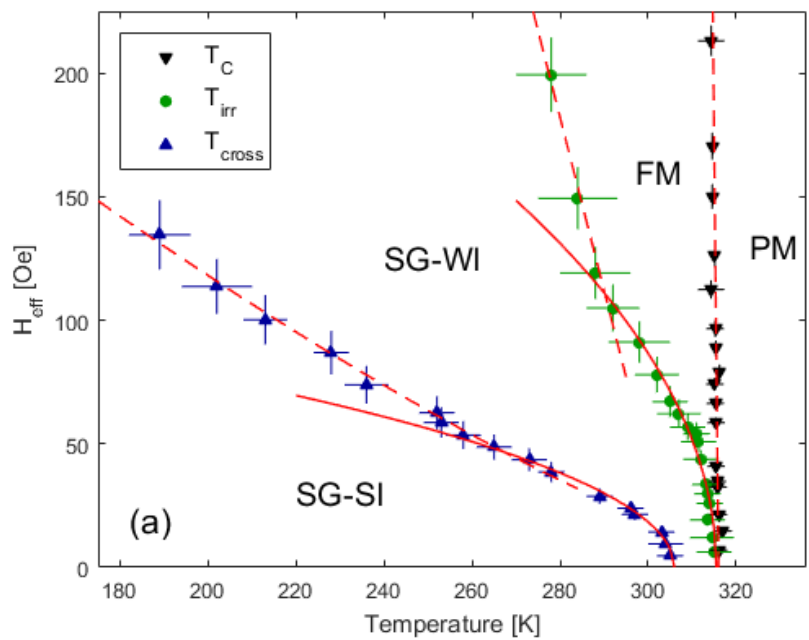

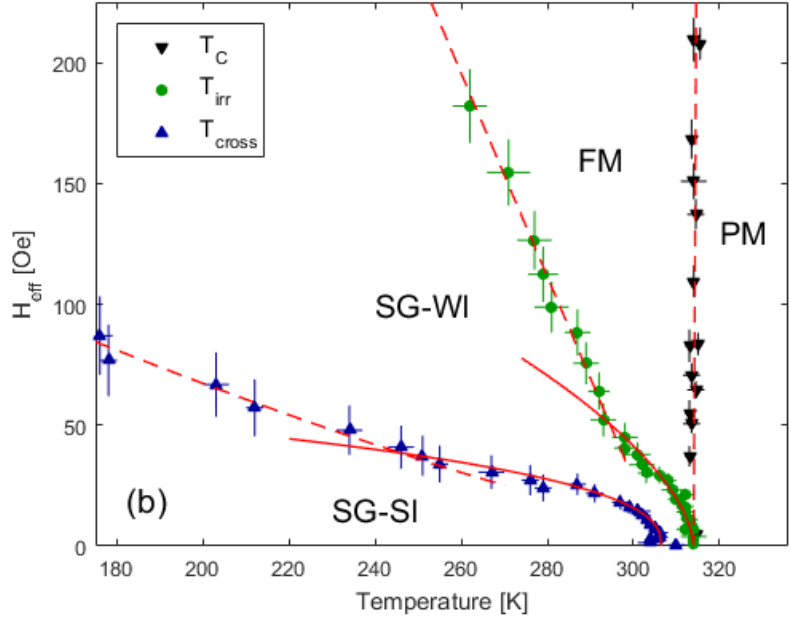

Fig. 2 Magnetic phase diagrams in the *H-T* plane for the investigated samples: *x*=33.5 (a), and *x*=34.1 (b). The dashed lines in the case of $T_C$ –values are to guide the eye, while other lines (solid and dashed) represent the best-fits to the corresponding data in terms of eq. (1). *PM* stands for a paramagnetic phase, *FM* for the ferromagnetic one, *SG-WI* marks the spin-glass state with a weak irreversibility, while *SG-SI* the one with a strong irreversibility.

The re-entrant character of magnetism in the studied samples is evident. As expected, an increase of the external magnetic field enlarges the ferromagnetic (*FM*) range but the actual enlargement sample's dependent. The $T_C$-values are *H*-independent which is indicative of a true phase transition (*PM*→*FM*). On the other hand, both $T_{irr}$ and $T_{cross}$ show a conspicuous decrease with the effective *H* which, however, for both temperatures changes its character from a non-linear to a linear (at $H\approx$ 50 Oe for *x*=33.5, and at $H\approx$ 40 Oe for *x*=34.1). To get a quantitative insight into the effect of *H* the data were analyzed in terms of eq. (1). The best-fit values of $\varphi$ are given in Table 1.

Table 1. Values of $\varphi$ as obtained for the studied samples by fitting eq. (1) to $T_{irr}$ and $T_{cross}$ data.

| x [at. % V] | 33.5 | | 34.1 | |
| --- | --- | --- | --- | --- |
| T | $\varphi$ | $H_{eff}$ [Oe] | $\varphi$ | $H_{eff}$ [Oe] |
| $T_{irr}$ | 2.0 | <~100 | 1.7 | <~50 |
| | 1.0 | >~100 | 1.0 | >~50 |
| $T_{cross}$ | 2.0 | <~50 | 2.5 | <~40 |
| | 0.75 | >~50 | 0.75 | >~40 |

It follows that for the sample with *x*=33.5 the decrease in both temperatures is in line with the *GT*-prediction i.e. φ=2 in low-field ranges (<~100 Oe for $T_{irr}$, and <~50 Oe for $T_{cross}$). In the high-field ranges, $T_{irr}$ decreases linearly with *H,* while $T_{cross}$ dependence reveals the value of φ=0.75. In other words, for both temperatures there is a critical value of *H* above which the effect of *H* on the temperature is weaker. Generally, similar behavior is observed for the sample with *x*=34.1 i.e. there exists also the critical magnetic field above which the values of φ are significantly different. Noteworthy, in the latter case the values of the critical field are much lower (especially for $T_{irr}$), while the corresponding values of φ are comparable. To get a deeper insight into the heterogeneous structure of the magnetic ground state of the investigated samples, an temperature-integrated difference between the FC and ZFC magnetization curves, <*ΔM*>, was calculated for each value of *H* based on the following formula:

$$<\Delta M> = \int_{T_1}^{T_2} (M_{FC} - M_{ZFC}) dT \quad (2)$$

Where $T_1$ = 80K and $T_2$=360K . Figure 3 illustrates a dependence of <*ΔM*> on the applied field, *H.* Its character is similar for both samples. Interestingly, in small fields (less than ~50 Oe) a steep increase of <*ΔM*> is observed, next a maximum occurs at $H \approx 50$ or 60 Oe followed by a moderate decrease. However, the two plots differ significantly as far as the values of <*ΔM*> are concerned viz. those for *x*=34.1. are significantly higher. If one assumes that <*ΔM*> can be taken as a measure of a degree of magnetic irreversibility than the results shown in Fig. 3 give a clear evidence that in the sample with the higher V content magnetic moments in the ground magnetic state are more irreversible (or spin-glassy) than the ones in the samples with a slightly lower content of V. In other words, the magnetic moments in the $Fe_{65.9}V_{34.1}$ sample in its ground state are more frozen than the ones in the $Fe_{66.5}V_{33.5}$ one. Noteworthy, the fields at which the maxima occur correspond fairly well to the ones at which the characteristic lines in the H-T plane have anomalies. The values of <*ΔM*>$_H$= ∫*ΔM·dH* can be regarded as energy of the irreversibility integrated over the temperatures, in which the irreversibility occurs viz. *ΔT*=<$T_{irr}$>-80K where <$T_{irr}$> is an average temperature of the irreversibility (304.9 K for *x*=33.5, and 307.4K for *x*=34.1). Dividing <*ΔM*>$_H$ by *ΔT* one gets a specific energy of irreversibility. Its value is 4.0 and 5.9 mJ/kg for x=33.5 and x=34.1, respectively. These figures again show a significant difference between the two samples as far as the magnetic ground state is concerned.

In summary, by measuring magnetization curves versus temperature in zero-field cooled and field could regimes and in external magnetic field of various amplitudes, magnetic phase diagrams in the *H-T* plane have been designed for two samples of the σ-$Fe_{100-x}V_x$ alloys (*x*=35.5 and *x*=34.1). They evidently testify to the re-entrant (PM→FM→SG) character of magnetism in the studied samples. The SG, being the ground magnetic, is magnetically heterogeneous viz. can be divided into a sub phase with a weak irreversibility (SG-WI) and the border line to the FM phase constituted by the values of $T_{irr}$, and the one with a strong irreversibility (SG-SI) and separated from the SG-WI sub phase by a line established from the $T_{cross}$-values. Both these lines show a cross-over behavior at a critical field i.e. change the rate of increase ($\varphi$) on decreasing temperature. Generally, in the low field and higher temperature range their behavior is in line with the GT-prediction ($\varphi=2$) while in the higher field and lower temperature domain the behavior is either linear ($\varphi=1$) or intermediate between the linear and that the AT-like ($\varphi=0.75$).

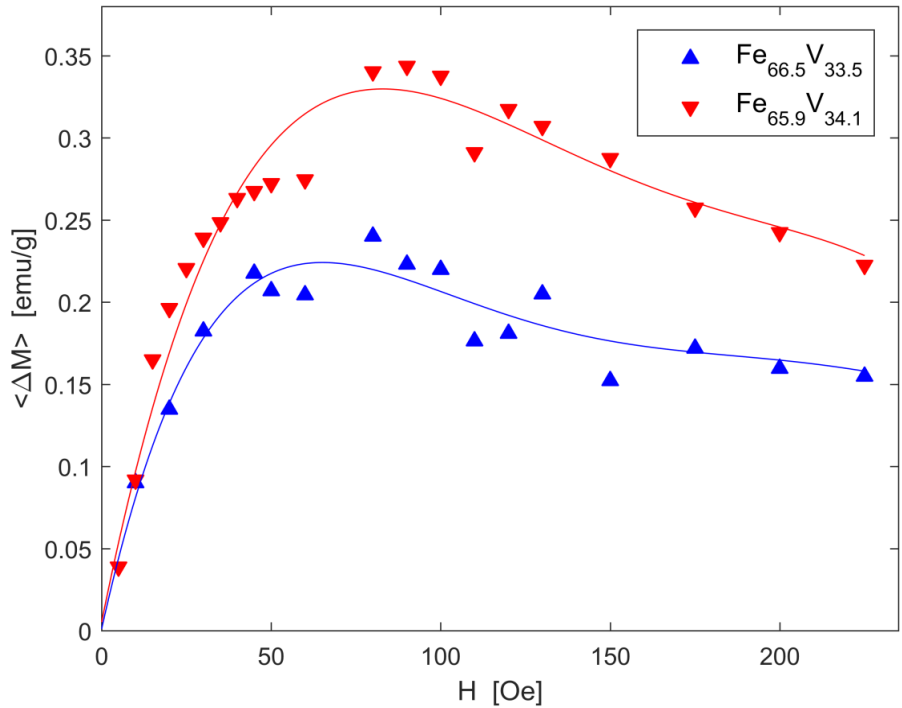

Fig. 3 Temperature integrated difference in the FC and ZFC magnetization curves as a function of an external magnetic field, *H*. The solid lines are guides to the eye.